# Magnetooptical properties of (Ga,Fe)N layers

J. Papierska[1], J.-G. Rousset[1], W. Pacuski[1], P. Kossacki[1], A. Golnik[1], M. Nawrocki[1],

J. A. Gaj[1], J. Suffczyński[1], I. Kowalik[2], W. Stefanowicz[2], M. Sawicki[2], T. Dietl[1,2],

A. Navarro-Quezada[3], B. Faina[3], Tian Li[3], A. Bonanni[3]

[1] *Faculty of Physics, University of Warsaw, Warsaw, Poland*

[2] *Institute of Physics Polish Academy of Sciences, Warsaw, Poland*

[3] *Johannes Kepler University of Linz, Austria*

Magnetooptical properties of (Ga,Mn)N layers containing various concentrations of Fe-rich nanocrystals embedded in paramagnetic (Ga,Fe)N layers are reported. Previous studies of such samples demonstrated that magnetization consists of a paramagnetic contribution due to substitutional diluted Fe ions as well as of ferromagnetic and antiferromagnetic components originating from Fe-rich nanocrystals, whose relative abundance can be controlled by the grow conditions. The nanocrystals are found to broaden and to reduce the magnitude of the excitonic features. However, the ferromagnetic contribution, clearly seen in SQUID magnetometry, is not revealed by magnetic circular dichroism (MCD). Possible reasons for differences in magnetic response determined by MCD and SQUID measurements are discussed.

## 1. Introduction

Previous magnetooptical studies on (Ga,Fe)N provided important information on the sp-d exchange interaction between free excitons and diluted localized Mn spins [1]. Doping of a semiconductor with transition metal ions above the solubility limit results in the formation of nanometer size, metal ion-rich regions differing either chemically or crystallographicaly from a host matrix [2]. Such decomposed systems offer several novel functionalities [2,3].

The aim of the present work is to investigate possible magnetooptical signatures originating from Fe-rich nanocrystals aggregating in (Ga,Fe)N films [4,5]. A contribution originating from ferromagnetic nanocrystals to magneto-optical Kerr effect (MOKE) and magnetic circular dichroism (MCD) has already been found for decomposed (Ga,Mn)As [6] and (Zn,Cr)Te [7].

Since transmission electron microscopy (TEM) measurements show that Fe nanocrystals in (Ga,Fe)N segregate towards the layer surface [3], magneto-reflectivity methods probing mainly the surface region of a sample, namely MCD MOKE, are chosen for the present study. In order to gain comprehensive information on the magnetooptical properties of Fe-rich nanocrystals, the obtained data are compared to the results of magnetometry characterization performed on the same samples.

## 2. Samples

Two (Ga,Fe)N layers grown by metalorganic vapor phase epitaxy (MOVPE) on a 1 μm thick GaN buffer deposited on the $Al_2O_3$ substrate are studied. The layers have similar concentrations of isolated Fe ions in paramagnetic phase (layer #1: $2\times10^{19}$ cm$^{-3}$ and #2: $3\times10^{19}$ cm$^{-3}$). Due to suitably adjusted growth conditions [2,3] layer #1 contains additionally Fe-rich nanocrystals with an overall Fe concentration determined by magnetometry to be $3\times10^{19}$ cm$^{-3}$ [3].

Pieces of both samples have been additionally sputtered from the surface side with Ar ions. Magnetometry measurements show that the sputtering reduces the concentration of the Fe-rich nanocrystals in #1 to $0.6\times10^{19}$ cm$^{-3}$, while the concentration of paramagnetic Fe ions in both layers remains unaffected.

## 3. Experiment

The spectroscopy measurements are performed in magnetic field up to 7 T applied in the Faraday configuration at a temperature of 2 K. In reflectivity measurement the samples are illuminated by a Xe lamp, with the incident beam parallel to the $c$-axis of the sample (spot diameter on the sample surface ~0.5 mm). Two circular polarizations of the light, σ+ and σ-, are detected and the MCD is determined according to the formula:

$$MCD = \frac{I_{\sigma+} - I_{\sigma-}}{I_{\sigma+} + I_{\sigma-}}$$

where $I_{\sigma+}$ ($I_{\sigma-}$) encodes the intensity of the signal in the σ+ (σ-) polarization.

In the MOKE measurements, the incident beam is linearly polarized and the angle of polarization rotation after the near-normal reflection from the sample is determined as a function of the photon energy.

Magnetometry measurements are performed in the temperature range from 2 K to 300 K for two orientations (parallel and perpendicular) of the magnetic field with respect to the $c$-axis of the sample.

## 4. Results

### 4.1. Magnetospectroscopy

As seen in Fig. 1d, the optical transitions of the free $A$ and $B$ excitons in (Ga,Fe)N are well resolved in the reflectivity spectrum of sample #2 (containing only isolated Fe ions). In the case of sample #1 (isolated Fe ions and Fe-rich nanocrystals) excitonic transitions are much less intense, wider and they merge into a single line, as evident from Fig. 1a. This indicates that presence of the Fe nanocrystals introduces disorder and/or results in an increase of the carrier density in the structure.

As seen in Fig. 1a) and 1d), the excitons undergo Zeeman splitting in magnetic field. The MCD curves determined based on the reflectivity curves detected in two circular polarizations at $B$ = 3 T are shown in Fig. 1b) and 1e), while the curves resulting from the Kerr rotation angle as a function of photon energy are reported in Figs. 1c) and 1f). The amplitude of both MCD and Kerr rotation is much smaller for sample #1 than for sample #2.

The area under the MCD and Kerr rotation curves in the excitonic region is integrated to provide a measure of the magnetooptically determined magnetization. The obtained dependences of the MCD magnitude on the magnetic field for both samples are given in Fig. 2. As it is seen, the dependences are well described by the paramagnetic Brillouin function with no clear indication for hysteresis in the case of layer #1 (see the inset of Fig. 2). The dependence of the Kerr rotation (not shown) corroborates this finding. Magnetometry characterization shows that the sample #2 gives purely paramagnetic signal, while the response of sample #1 contains also a ferromagnetic component persisting up to the room temperature for both measurement configurations, *i.e.* a hysteresis loop with a coercive field of 0.05 T is observed. Since the ferromagnetic component of the signal from sample #1 has been attributed to nanocrystals and no ferromagnetic response is observed in MCD and MOKE, we conclude that the nanocrystals do not contribute to the magnetooptical signal in a sizable way. Instead, as demonstrated by Figs. 1 and 2, the presence of Fe-rich nanocrystals broadens and reduces the magnitude of excitonic features, and as a result reduces the magnitude of the magnetooptical signal from the (Ga,Fe)N layer.

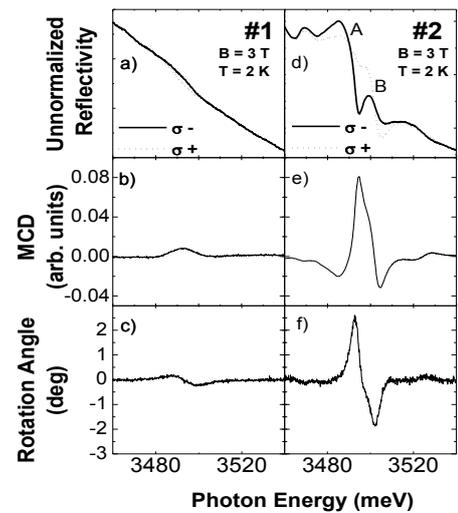

Fig. 1. (a, d) Reflectivity spectra taken at both circular polarizations of the light, (b, e) MCD and (c,f) Kerr rotation angle spectra for samples #1 and #2 as a function of photon energy at B = 3 T, T= 2 K.

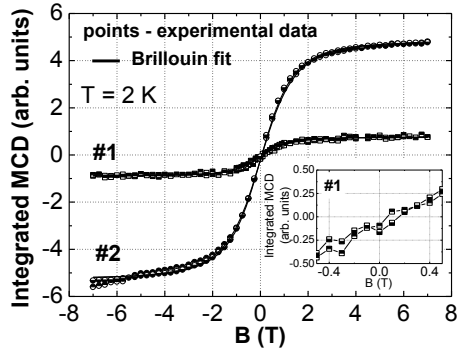

Fig. 2. Integrated MCD for samples #1 and #2 (points) together with the Brilluoin function fit (solid line) as a function of magnetic field. The magnetic field is varied subsequently in two opposite directions. Inset: close-up of the zero field region of the plot for sample #1.

### 4.2 Effects of Ar ion sputtering

As shown by magnetometry, Ar ion sputtering of the sample surface results in the vanishing of the ferromagnetic component of the signal from sample #1, while the response from sample #2 remains unaltered. Since Fe nanocrystals are located predominantly near the film surface this result confirms previous findings [3] that the ferromagnetic signal of the films comes from the nanocrystals.

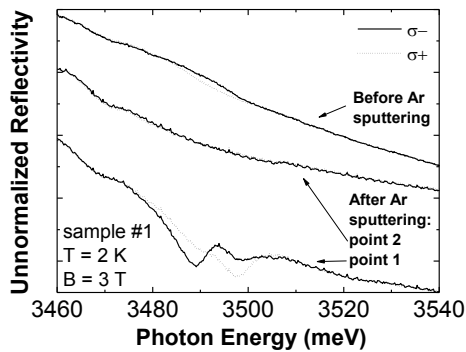

Fig.3 Reflectivity spectra from layer #1 collected before and after Ar sputtering at different points on the sample surface. A magnetooptical signal is recovered (point 1) or decreased (point 2) after sputtering depending on the region of the layer. Spectra shifted for clarity.

The intensity of the excitonic transitions seen in reflectivity spectra taken at different points of the sputtered samples differ, revealing a spatial inhomogeneity of the layers. In general, a decrease of the excitonic signal in the case of sample #2 is observed (not shown), and attributed to increased disorder. As seen in Fig. 3, sputtering leads to either increase or decrease of the magnetooptical signal of the sample #1. The increase is attributed to the effect of the lowering of the Fe nanocrystals density revealed by magnetometry. This finding goes in line with the observation of larger magnetooptical signal from unsputtered layer #2 as compared to #1.

### 5. Conclusions

The magnitude of excitonic transitions in (Ga,Fe)N and the resulting magnetooptical signal in magnetic field are found to be anticorrelated with the concentration of Fe-rich nanocrystals. No proof of ferromagnetic contribution to the magnetooptical signal related to Fe-rich nanocrystals is found. This suggests that either of the three factors: (i) nanocrystals distribution, (ii) their density or (iii) coupling strength to excitons does not result in a sizable magnetoptical phenomena. Also, the increased signal to noise ratio resulting from lowered magnitude of the excitonic signal in the case of the layer with high concentration of nanocrystals is an obstacle for a quantitative analysis of the magnetooptical effects and may preclude the observation of hysteresis loop.

One of the solutions would be to investigate samples with Fe-rich nanocrystals distributed homogeneously in the layer volume and not - as in the present case - predominantly at the surface proximity. Alternatively, one could turn to photoluminescence experiments and enhance the emission intensity lowered by the presence of Fe-rich nanocrystals by exploiting plasmonic effects [8].

### Acknowledgments

The work was supported by FunDMS Advanced Grant of ERC within the Ideas 7th FP of EC, NCBiR project LIDER, MNiSW project Juventus Plus, and InTechFun (Grant No. POIG.01.03.01-00-159/08).